\documentclass[twocolumn,showpacs,aps,amssymb,floatfix,prd,amsmath,preprintnumbers,10pt]{revtex4} 
\usepackage{graphicx}  
\usepackage{dcolumn,color}   
\usepackage{bm}
\begin{document}
\def \reals{{\mathbb R}}
\def\be{\begin{equation}}
\def\ee{\end{equation}}
\def\bea{\begin{eqnarray}}
\def\eea{\end{eqnarray}}
\def\nn{\nonumber}
\def\th{\theta}
\def\ph{\phi}
\def\lt{\left}
\def\rt{\right}
\def\degree{\mathop{\rm {{}^\circ}}}
\input epsf.tex
\title{Distortions of  images of Schwarzschild lensing}\thanks{ This paper is  dedicated to the memory of my {\em guru} 
 Prof.~ P.~C.~Vaidya. (In Sanskrit, {\em gu} means darkness and {\em ru} means one who dispels.)}

\author{K. S. Virbhadra}
    \email[Email address : ]{shwetket@yahoo.com}
    \affiliation{Mathematics Department, Drexel University, 33rd and Market Streets, Philadelphia, Pennsylvania 19104, USA}

\begin{abstract}
We model the supermassive dark object $M87^*$ as a Schwarzschild lens and study the variations in tangential, radial, and total (the product of tangential and radial) magnifications of images (primary, secondary, and relativistic) against the changes in angular source position and the ratio of lens-source to the observer-source distance. Further, we study the behavior of partial derivatives (with respect to the angular source position) of total magnifications of images against the angular source position. Finally, we model supermassive dark objects at centers of 40  galaxies as Schwarzschild lenses and study the variations in tangential, radial, and total magnifications of images against  the change in the ratio of mass of the lens to its distance. These studies yield  many nonintuitive results which are likely to be significant for next generation Event Horizon Telescope  observations. We {\em hypothesize} that there exists a  distortion parameter such that their signed sum of all images of  singular gravitational lensing of a source identically vanishes. We test this with images of Schwarzschild lensing in weak and  strong gravitational fields and find that this esthetically appealing hypothesis succeeds with flying colors.
\end{abstract}

\pacs{95.30.sf, 04.20.Dw, 04.70.Bw, 98.62.Sb }

\keywords{Gravitational lensing, black holes, naked singularities, relativistic images, and distortion}

\maketitle

.
\section{\label{sec:Intro}Introduction}

The deflection of light in the gravitational field of a massive object was discussed even before the discovery of Einstein's general theory of relativity, notably by Newton, Michell, Cavendish, Laplace, Soldner, and Einstein himself \cite{Book1,Book2,NB96,Book3}. Later, after the advent of general relativity, Einstein, using his theory of general relativity, obtained total light deflection of a light ray tangentially grazing the surface of the Sun. His result was twice the Newtonian value and  was supported by observation during the total solar eclipse in 1919. 

The spectacular detectable phenomena resulting from the deflection of electromagnetic or gravitational radiation by a spacetime is referred to as {\em gravitational lensing}\cite{ModifiedDef}.
The basic  theory of lensing was developed by Eddington, Chwolson, Einstein, Liebes, Klimov, Refsdal, Bourassa and Kantowski, and others (see in  \cite{Book1} and references therein.) In 1979, Walsh, Carswell, and Weymann \cite{WCW79} observed  twin images of QSO 0957+561 A, B that were separated by approximately  5.7 {\em arcsec}.  Thereafter, Lynds and Petrosian \cite{LP86}, in 1986,  and Soucail {\it et al.} \cite{Sou87},  in 1987,  observed giant luminous arcs which were unraveled as distorted images of distant galaxies by Paczynski \cite{Pac87}. Hewitt {\it et al.} \cite{Hew88}, in 1988,  observed the first Einstein ring. These observations made gravitational lensing one of the hottest research topics in astrophysics and numerous gravitational lensing cases have been observed by now. Around two  decades before the first GL (gravitational lensing) was observed by Walsh {\it et al.}, Darwin \cite{Dar59}, in 1959,  carried out basic studies of the gravitational lensing due to light deflection in the vicinity of the photon sphere of an ultracompact Schwarzschild massive object. He obtained an elegant formula for the Einstein bending angle:
\begin{equation}
\hat{\alpha}\lt(r_o\rt) = 2 \ln\lt[   \frac{36M(2-\sqrt3)}{r_o-3M}\rt] - \pi \text{,}
\label{DarwinAlpha}
\end{equation} 
where $r_o$ and $M$ are, respectively, the closest distance of approach of the light ray  and the Schwarzschild mass. He further showed that the images are too demagnified to be observed and termed those {\em ghosts} probably because those were not observable. Despite the theoretical elegance  of results obtained by Darwin, the research on gravitational lensing in a  strong gravitational field remained  almost  abeyant     for around 40 years for possibly two reasons: Images were incredibly demagnified and there was no adequate gravitational lens equation to study GL in a very strong gravitational field.

 Being unaware of Darwin's work, we \cite{VE00},  in 2000, initiated  a research on this topic. We obtained a new gravitational lens equation that allows arbitrary  light deflection angles (very small through very large). We  modeled the Galactic supermassive ``black hole" as a Schwarzschild lens and obtained angular positions of primary-secondary as well as {\em relativistic images} (deflection angle $\hat{\alpha} > 3 \pi/2$)  and their magnifications. Like Darwin,  we also found that  relativistic images (that he called ghosts) are  very demagnified. However, the new lens equation being  capable of studying GL in a very strong gravitational field resurrected the  strong field gravitational lensing studies. Perlick \cite{Per06} called this new lens equation  an {\em almost exact lens equation}. A large number of research papers on strong gravitational field lensing due to   black holes \cite{CVE01,Perlick04,Perlick22,Atamurotov21,Hsieh21,Tsukamoto21,Chagoya21,Alawadi21,Kumar20,Bisnovatyi17,Tsukamoto17,Amarilla12,Frittelli00,Chowdhuri21} and exotic objects such as naked singularities, wormholes, and boson stars \cite{VNC98,VE02,VK08,Gyulchev20,Gyulchev08,Jusufi19,Sahu13,Nakajima14,DeAndrea14,Bellorin14,Nandi06,Dey08,Kuhfittig14,Schunck06,Gao19}  have been  published (see also references therein.)

In 2009, we \cite{Vir09} came back to this subject and carried out a comprehensive study of Schwarzschild lensing in a weak as well as  the strong gravitational field in the vicinity of the photon sphere. We obtained important results not only for relativistic images but also for primary-secondary images which were thought  to be completely understood. We   list a few of those results in brief: 
(i) We obtained a  formula  for  computing  masses of compact objects with astounding accuracy. Distances and angular source position play no (extremely insignificant) role in this formula. [See Eq. ($19$) in \cite{Vir09}.]
(ii) The angular separations between any two relativistic images are extremely insensitive to changes in the lens-source distance and angular source position. Thus, having  the mass of the compact object and separation between angular positions between two relativistic images,  we can compute a very accurate value for the distance of the compact massive object.      
(iii) The dependence of (absolute) magnification ratios of relativistic images of the same order on the potential (the ratio of the mass of the lens  to the lens-observer distance) is insignificantly  small. Therefore, the measurements of the flux ratio would give a very accurate value of the lens-source distance.  
(iv) We showed that the time delays of primary images are always (for any angular source position) smaller for nearer sources for the otherwise same situation. This is obviously a counterintuitive result. We also explained the reason for this long prevailing misconception. Usually, differential time delays are measured. However, this conceptually fascinating  counterintuitive result is  also measurable.  Rafikov and Lai \cite{RL06}   gave  a method to measure time delay as well. For recent papers on the gravitational lensing in strong gravitational fields, see \cite{Ish16,Ish17,Ono17,Taki20,Bog22,Vir22,AdlerVir22} and references therein.
    
 Despite the wonderful implications of  relativistic images for astrophysics (as these could bestow a  powerful  means to unveil the secrets of the universe with astounding accuracy), these are not  observed. To this end, after a long  period of technical and theoretical developments, the event horizon telescope (EHT)  --- an international collaboration from many countries and institutions was launched in 2009. The EHT was composed of radio telescopes around the world to produce a giant high sensitivity and resolution virtual telescope. In 2017 the EHT carried out observations of $M87^*$  over 4 days (April 5-11) at  approximately 1.3 mm with unprecedented angular resolution. They  surprised the world by releasing the first image on April 10, 2019 and  published their landmark results in a series of six papers\cite{EHT1,EHT2,EHT3,EHT4,EHT5,EHT6}. The bright region around the silhouette seems to be of secondary, relativistic,  and {\em orphaned images}. (The images due to gravitational mirroring, also called retrolensing, of a source do not have primary-secondary images as their ``parents"  and due to this reason we called those  orphaned images, or simply orphans \cite{Vir09}.) The  relativistic images are easy to be differentiated  from orphans because these appear along with the primary and secondary images. The present EHT is not capable of  resolving  relativistic images from secondary as well as orphaned images  and therefore  we are not sure whether the EHT  observed relativistic images. However,  the next generation Event Horizon Telescope (ngEHT)\cite{ngEHT}    is very likely to resolve secondary, relativistic, and orphaned images and also take necessary measurements. 
    
 In order to observe and analyze relativistic images, we need to study characteristics  of these images in detail. This is the main aim of this paper. We first model the $M87^*$ as the Schwarzschild lens and study variations in tangential, radial, and total magnifications of images with respect to the change in  the angular source positions (keeping lens-source distance fixed). Then, we study  derivatives of total  magnifications (with respect to angular source position)  as the angular source position increases.   We further study variations in  magnifications   with respect to the change in distance parameter (the ratio of lens-source to observer-source distances) keeping the angular source position fixed.  Last, we model the supermassive compact objects of 40 galaxies as the Schwarzschild lenses and study the variations of magnifications with respect to the ratio of the mass to the distance of the lens, keeping the  ratio of lens-source to the  observer-source distances and angular source position  fixed. Last, but not  least important, we define a novel {\em distortion parameter} of images such  that   the sum of signed (not absolute) distortions of all images is zero. With numerical computations, we demonstrate that   the distortion parameter has this characteristic with  high accuracy.  The inclusion of this distortion parameter in the theory of gravitational lensing could be very helpful in identifying images of the same source and the order as well as searching for missing image(s).

 This paper is arranged as follows. In Sec. II, we give lens equation, magnifications, and definition  of  distortion parameter. In Sec. III, we carry out computations and present results. In Sec. IV, we give a summary and discuss the results. We use geometrized units (the universal gravitational constant $G = 1$ and the speed of light in vacuum $c=1$)  and therefore the mass  $M \equiv M G/c^2$. We  use {\em Mathematica} \cite{Math12} for computations.
             
\section{\label{sec:LE}Lens Equation, Magnification, and Distortion}
In order to study gravitational lensing due to light deflection in weak as well as strong gravitational fields (such as in the vicinity of photon surfaces of compact massive objects), we obtained a novel lens equation which is expressed as \cite{VE00}
\be
\tan\beta =  \tan\theta - \alpha ,
\label{LensEqn}
\ee
where
\be
\alpha =
    {\cal D}  \lt[\tan\theta + \tan\lt(\hat{\alpha} - \theta\rt)\rt] 
\label{Alpha}
\ee
with 
\be
{\cal D} = \frac{D_{ds}}{D_s} \text{.}
\ee
Symbols $\beta$ and  $\theta$, respectively, stand  for the angular positions of the unlensed source and image.  $\hat{\alpha}$ represents the Einstein bending angle of the light ray. The impact parameter 
\be
J = D_d \sin\theta \text{.}
\ee
The symbols $D_d, D_{ds}$, and $D_s$ denote, respectively, the observer-lens, lens-source, and observer-source angular diameter distances. The subscript $d$ stands for the deflector (lens).  The values of the  dimensionless distance parameter, $D$, lie in the interval $\left(0, 1\right)$. However, its values should not be chosen too close to zero  (sources not too close to photon surfaces) in order for the lens equation to work well.

The magnification of an image is defined as the ratio of the flux of the image to the flux of the unlensed source.  However,  according to  Liouville's theorem,  the surface brightness is conserved in light deflection and therefore  this ratio turns out to be the ratio of solid angles of the image and of the unlensed source. Thus, the total magnification of an image of a circularly symmetric  gravitational lensing is 
\be
\mu = \mu_t \mu_r \text{,}
\label{Mu}
\ee
where the  tangential $\mu_t$  and radial  $\mu_r$ magnifications are, respectively,  given  by
\be
\mu_r = \lt(\frac{d\beta}{d\theta}\rt)^{-1} ~ ~ ~  \text{and} ~ ~ ~ 
\mu_t = \lt(\frac{\sin{\beta}}{\sin{\theta}}\rt)^{-1} \text{.}
\label{MutMur}
\ee
The sign of the magnification of an image determines parity of the image:  positive parity for $\mu > 0$, negative parity for $\mu < 0$, and zero parity for the images formed when  the angular source position $\beta = 0$. However, $\beta = 0$ does not always give images \cite{VNC98, VE02,VK08}. If the tangential magnification $\mu_t$ of an image is negative,  then we define absolute tangential magnification (but may  be simply  called tangential magnification) $|\mu_t|$ for plotting and analysis of results. The same convention applies to  radial and total magnifications.

In this paper, we study  gravitational lensing by static spherically symmetric compact objects. The exterior gravitational field of such objects is given by the Schwarzschild spacetime described by the following line element:
\begin{eqnarray}
ds^2&=&\lt(1-\frac{2M}{r}\rt)dt^2- \lt(1-\frac{2M}{r}\rt)^{-1} dr^2 \nonumber \\
      &-&r^2\lt(d\vartheta^2+\sin^2 \vartheta d\phi^2\rt) \text{,}
   \label{SchMetric}
\end{eqnarray}
where the real constant parameter $M$ is the Schwarzschild mass. The deflection angle $\hat{\alpha}$ and the impact parameter $J$ of a light ray with the closest distance of approach $r_o$ are given, respectively, by \cite{Wei72}
\be
\hat{\alpha}\lt(r_o\rt) = 2 \  {\int_{r_o}}^{\infty}
 \frac{dr}{r \  \sqrt{\lt(\frac{r}{r_o}\rt)^2  \lt(1-\frac{2M}{r_o}\rt)
-\lt(1-\frac{2M}{r}\rt)}   } - \pi 
\label{AlphaHatR0}
\ee
and 
\be
J\left(r_o\right) = r_o \lt(1-\frac{2M}{r_o}\rt)^{-1/2} .
\label{ImpParaR0}
\ee
Like in our previous papers \cite{VNC98,VE00}, we introduce radial distance in terms of the Schwarzschild radius $2M$,
\be
\rho = \frac{r}{2M} , ~ ~ ~
\rho_o = \frac{r_o}{2M} , 
\label{XX0}
\ee
and write
\begin{equation}
  \hat{\alpha}\lt(\rho_o\rt) = 2 \  {\int_{\rho_o}}^{\infty}
 \frac{d\rho}{\rho  \  \sqrt{\lt(\frac{\rho}{\rho_o}\rt)^2 
 \lt(1-\frac{1}{\rho_o}\rt)
-\lt(1-\frac{1}{\rho}\rt)}} - \pi 
   \label{AlphaHatRho0}
\end{equation}
and
\begin{equation}
  J\lt(\rho_o\rt) = 2M  \rho_o \lt(1-\frac{1}{\rho_o}\rt)^{-1/2}.
     \label{ImpParaRho0}
\end{equation}
In order to compute magnifications of  images, we need derivative of the deflection angle $\hat{\alpha}$ with respect to the angular position of the image which is given by [see Eq. (32)  in \cite{VNC98}]
\be
\frac{d\hat{\alpha}}{d\theta} =  \hat{\alpha}'\lt(\rho_o\rt) \frac{d\rho_o}{d\theta} ,
    \label{DAlphaByDTheta}
\ee
\begin{widetext}
\be
\hat{\alpha}'\lt(\rho_o\rt) = \frac{3-2\rho_o}{{\rho_o}^2\lt(1-\frac{1}{\rho_o}\rt)}
{\int_{\rho_o}}^{\infty}
 \frac{\lt(4 \rho - 3\rt) d\rho}
{\lt(3 - 2 \rho\rt)^2 \  \rho \   \sqrt{\lt(\frac{\rho}{\rho_o}\rt)^2 
 \lt(1-\frac{1}{\rho_o}\rt)
-\lt(1-\frac{1}{\rho}\rt)}} 
 \label{DAlphaHatByRho0}
\ee
\end{widetext}
and  
\begin{equation}
 \frac{d\rho_o}{d\theta} = 
     \frac{  \rho_o \lt(1-\frac{1}{\rho_o}\rt)^{3/2}
        \sqrt{1-\lt(\frac{2M}{D_{d}}\rt)^2 {\rho_o}^2 \lt(1-\frac{1}{\rho_o}\rt)^{-1}}}
        {\frac{M}{D_{d}} \lt(2\rho_o-3\rt)} \text{.}
      \label{DRho0ByDTheta}
\end{equation}
Different images of the same source are usually identified by the similar spectra, same flux ratio in the optical as well as radio wave band, and knots in the different images. However, there is no way to know if there is/are any missing image(s) of the same source for whatever reasons. Inspired by this  problem, we hypothesize the following: There exists a {\em distortion parameter} such that the signed (not absolute) sum of distortions of all images of a realistic and singular gravitational lensing of a source is identically zero. (We assume that no image is occulted.)
In order to find a parameter like this, we define a distortion  parameter as 
\be
\Delta = \frac{\mu_t}{\mu_r} \text{.} 
 \label{Delta}
\ee
The signed sum of distortions of all images of a given source 
\be
{\Delta}_{sum} = \sum_{i=1}^{k} \Delta_i \text{,}
  \label{SumDistortions}
\ee
where $k$ is the total number of images. We also define a  {\em logarithmic distortion parameter} of an image
\be
 \delta = \log_{10} \big|\frac{\mu_t}{\mu_r}\big|
\label{delta}
 \ee 
for the convenience in plotting. We use the subscripts p and s for the primary and secondary images and subscripts $p_1$ and $p_2$ for the relativistic  images of the first and second orders respectively, on the primary image side. Similarly, we use  subscripts $s_1$ and $s_2$ for relativistic images on the secondary image side. We defined a (signed) distortion parameter $\Delta$ with the aim that sum of  distortions of all images be zero. In order to test whether images of the same order have the same absolute distortions, we now define {\em   percentage difference in distortions} of images of the same order as follows:
\bea
\mathbb{P}_{ps}  &=&     \frac{\Delta_p +\Delta_s}{\Delta_p}\times 100 \text{,}  \nn \\
\mathbb{P}_{1p1s}  &=&   \frac{\Delta_{1p} +\Delta_{1s}}{\Delta_{1p}}\times 100 \text{,}  \nn \\
\mathbb{P}_{2p2s}  &=&   \frac{\Delta_{2p} +\Delta_{2s}}{\Delta_{2p}}\times 100 \text{,} 
\label{PercentageDiff}
\eea
where the  subscript $ps$ stands for the primary-secondary pair, and $1p1s$ and $2p2s$, respectively, for the first-  and second-order relativistic images pairs.
As the primary image and relativistic images on the primary image side have positive parity in Schwarzschild lensing, the distortions of these images (i.e., $\Delta_p$, $\Delta_{1p}$, and  $\Delta_{2p}$) are also positive.  Similarly, distortions of secondary  image and relativistic images on the secondary image side  (i.e.,  $\Delta_s$, $\Delta_{1s}$, and  $\Delta_{2s}$) are  negative and due to this reason these appear  with a positive sign in the numerators of the absolute percentage difference formulae given above. 
\section{\label{sec:Mag}Computations and Results}

\begin{figure*}[tbh]
\centerline{ \epsfxsize 17.3cm
   \epsfbox{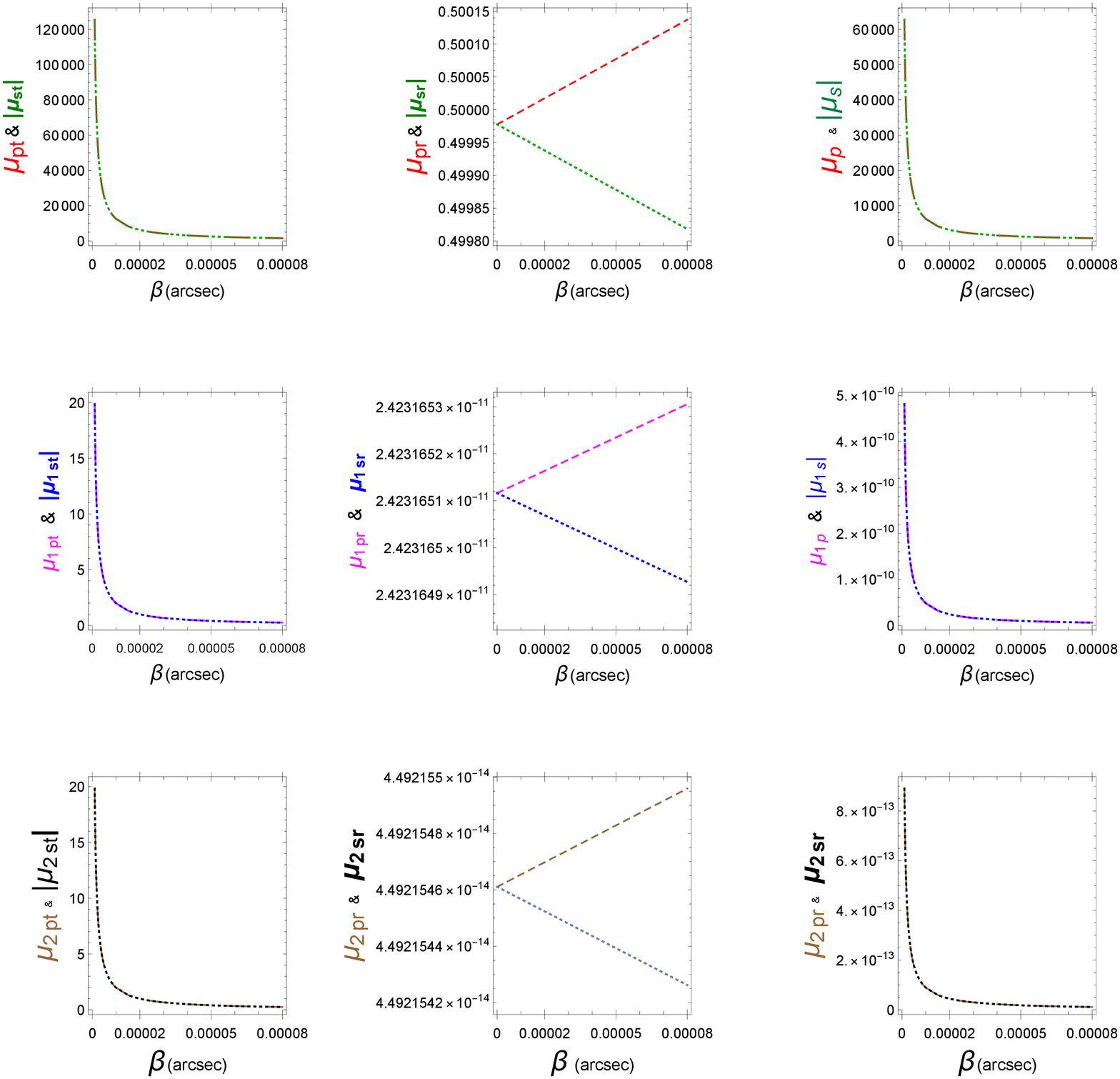}}
 \caption[ ]{
{\em First row}:  the tangential magnification of the primary image $\mu_{pt}$ (red dashed),  the absolute tangential magnification of the secondary image $\left|\mu_{st}\right|$ (green dotted), the radial magnification of the primary image $\mu_{pr}$ (red dashed),  the radial  magnification of the secondary image $\mu_{sr}$ (green dotted),
 the  total magnification of the primary image $\mu_p$ (red dashed),  and  the absolute total magnification of the secondary image $\left|\mu_s\right|$ (green dotted)
        are plotted against the angular source position $\beta$.
{\em Second and third rows}:  the same three magnifications are plotted against $\beta$ for the first (see the second row) and the second  (see the third row) order relativistic images. The subscripts $1p$ and $2p$ stand for
the first- and second-order relativistic images on the side of the primary image whereas subscripts $1s$ and $2s$  for
the first- and second-order relativistic images on the secondary image side. The colors of symbols and corresponding graphs are kept the same for graphs to be identified.
The  supermassive dark object (SMDO)  at the galactic center of M87 is modeled as the Schwarzschild lens, which has  mass $M= 6.5 \times 10^9 M_{\odot}$
and is situated at  the distance $D_d =  16.8$ {\em Mpc}  so that  $M/D_d \approx 1.84951\times  10^{-11}$.  The dimensionless parameter ${\cal D} = 0.005$.}
\label{fig1}
\end{figure*}
\begin{figure*}[tbh]
\centerline{ \epsfxsize 17.0cm
   \epsfbox{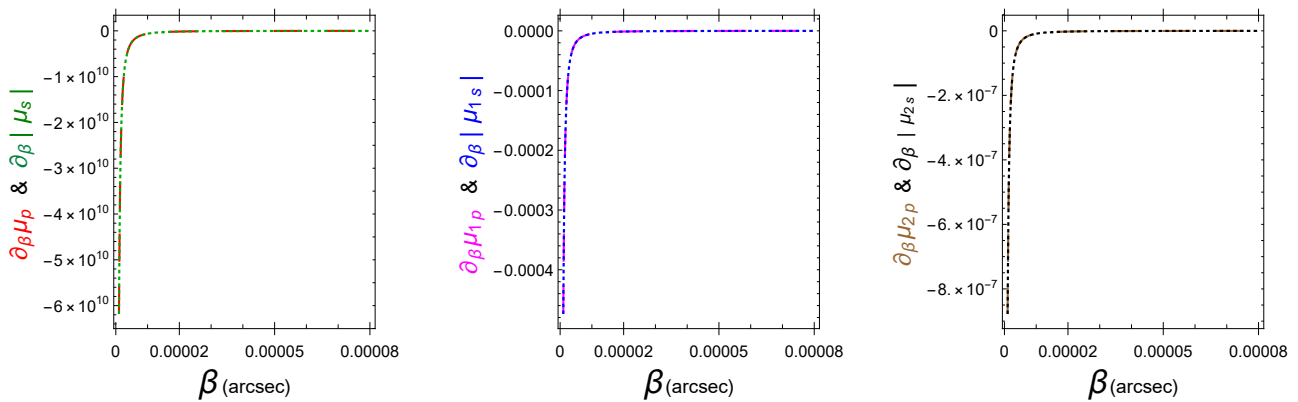}}
 \caption[ ]{
{\em Left}:  the partial derivatives   of total magnifications of  primary and secondary images, denoted respectively, by $\partial_\beta \mu_p$ (red dashed) and  $\partial_\beta \left|\mu_s\right|$ (green dotted)
            are plotted against the angular source position $\beta$.
{\em Middle}:  the partial derivatives   of total magnifications of the first order relativistic images on  primary and secondary sides, denoted respectively, by $\partial_\beta \mu_{1p}$
               (magenta dashed) and  $\partial_ \beta \left|\mu_{1s}\right|$ (blue dotted)   are plotted against $\beta$.
{\em Right}:  the partial derivatives   of total magnifications of the second-order relativistic images on  primary and secondary sides, denoted, respectively, by $\partial_\beta \mu_{2p}$
               (brown dashed) and  $\partial_ \beta \left|\mu_{2s}\right|$ (black dotted)   are plotted against the angular source position $\beta$.
The gravitational lens, as well as the parameter ${\cal D}$, are the same as for Fig. 1. Angular source positions are  expressed in {\em arcsec}.
 }
\label{fig2}
\end{figure*}

\begin{figure*}[tbh]
\centerline{ \epsfxsize 17.0cm
   \epsfbox{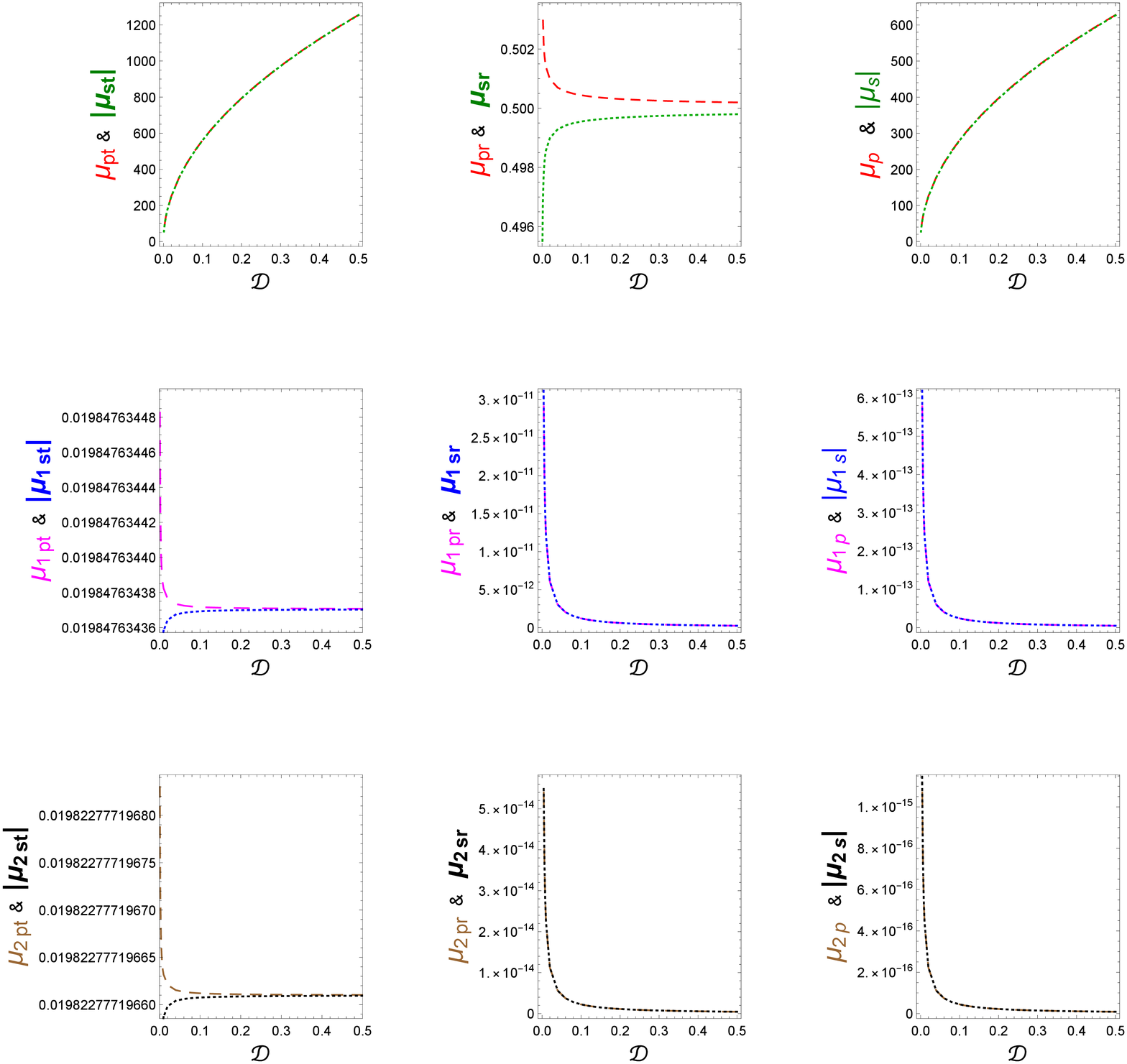}}
 \caption[ ]{
{\em First row}:  the tangential magnification of the primary image $\mu_{pt}$ (red dashed),  the absolute tangential magnification of the secondary image $\left|\mu_{st}\right|$ (green dotted),
                   the radial magnification of the primary image $\mu_{pr}$ (red dashed),  the               radial  magnification of the secondary image $\mu_{sr}$ (green dotted),
 the  total magnification of the primary image $\mu_{p}$ (red dashed),  and  the absolute total magnification of the secondary image $\left|\mu_{s}\right|$ (green dotted)
        are plotted against the parameter ${\cal D}$.
{\em Second and third rows}:  the same three magnifications are plotted against the parameter ${\cal D}$  for the first (see the second row) and the second  (see the third row) order relativistic images. The subscripts $1p$ and $2p$ stand for
the first- and second-order relativistic images on the side of the primary image whereas subscripts $1s$ and $2s$  for
the first- and second-order relativistic images on the secondary image side. The colors of symbols and graphs are kept the same for graphs to be identified.
The gravitational lens is the same as for Figs. 1 and 2. The angular source position $\beta = 1 mas$.
 }
\label{fig3}
\end{figure*}

\begin{figure*}[tbh]
\centerline{ \epsfxsize 17.0cm
   \epsfbox{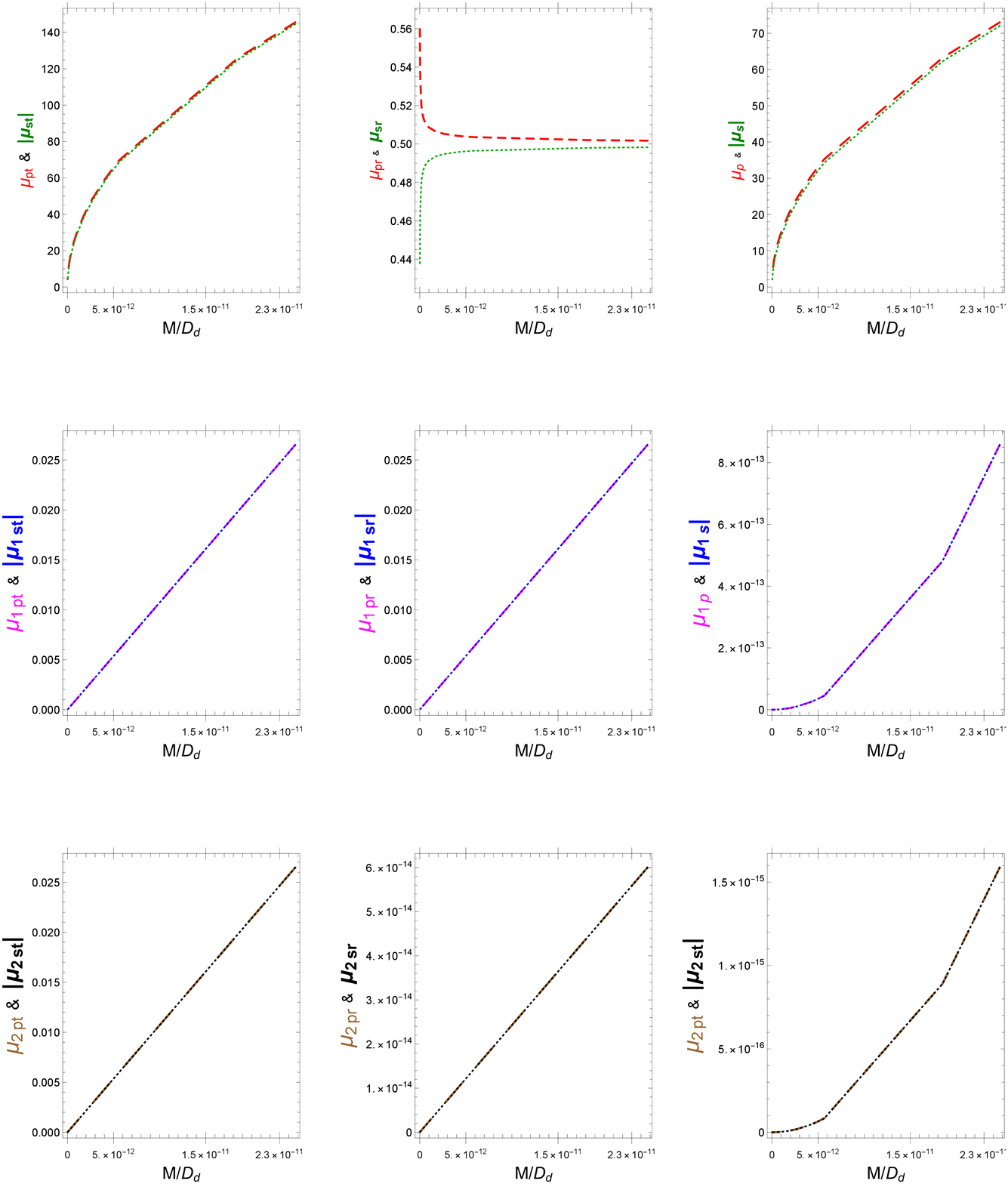}}
 \caption[ ]{
{\em First row}:  the tangential magnification of the primary image $\mu_{pt}$ (red dashed),  the absolute tangential magnification of the secondary image $\left|\mu_{st}\right|$ (green dotted),
                   the radial magnification of the primary image $\mu_{pr}$ (red dashed),  the               radial  magnification of the secondary image $\mu_{sr}$ (green dotted),
 the  total magnification of the primary image $\mu_{p}$ (red dashed),  and  the absolute total magnification of the secondary image $\left|\mu_{s}\right|$ (green dotted)
        are plotted against  $M/D_d$ (the ratio of the mass of the lens to the lens-observer distance).
{\em Second and third rows}:  the same three magnifications are plotted against  $M/D_d$  for the first (see the second row) and the second  (see the third row) order relativistic images. The subscripts $1p$ and $2p$ stand for
the first- and second-order relativistic images on the side of the primary image whereas subscripts $1s$ and $2s$  for
the first- and second- order relativistic images on the secondary image side. The colors of symbols and corresponding graphs are kept the same for graphs to be identified.
 The angular source position $\beta = 1 mas$ and  ${\cal D} = 0.005$.
}
\label{fig4}
\end{figure*}
\begin{figure*}[tbh]
\centerline{
     \epsfxsize 5.5cm  \epsfbox{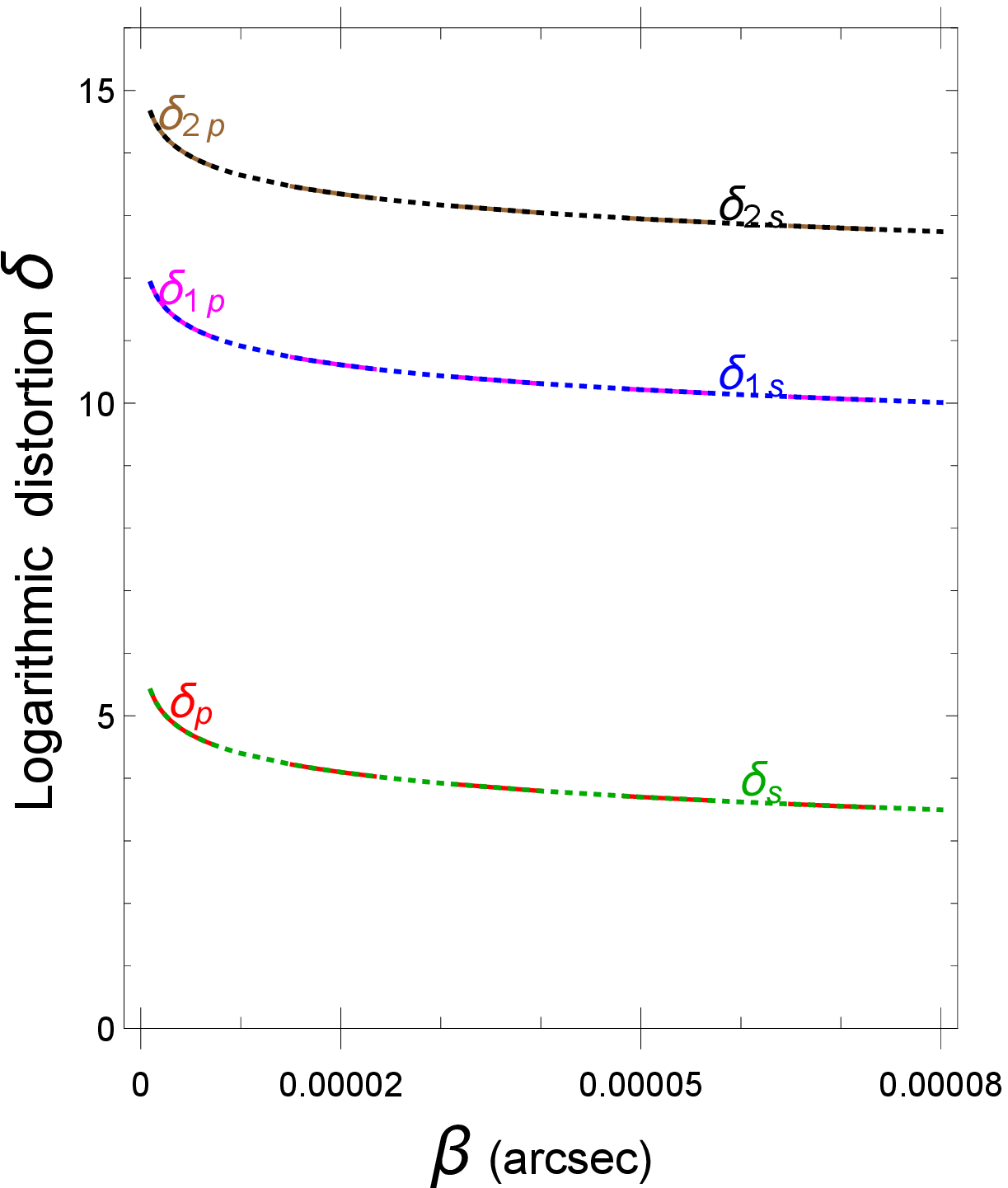}  \qquad      \epsfxsize 5.5cm  \epsfbox{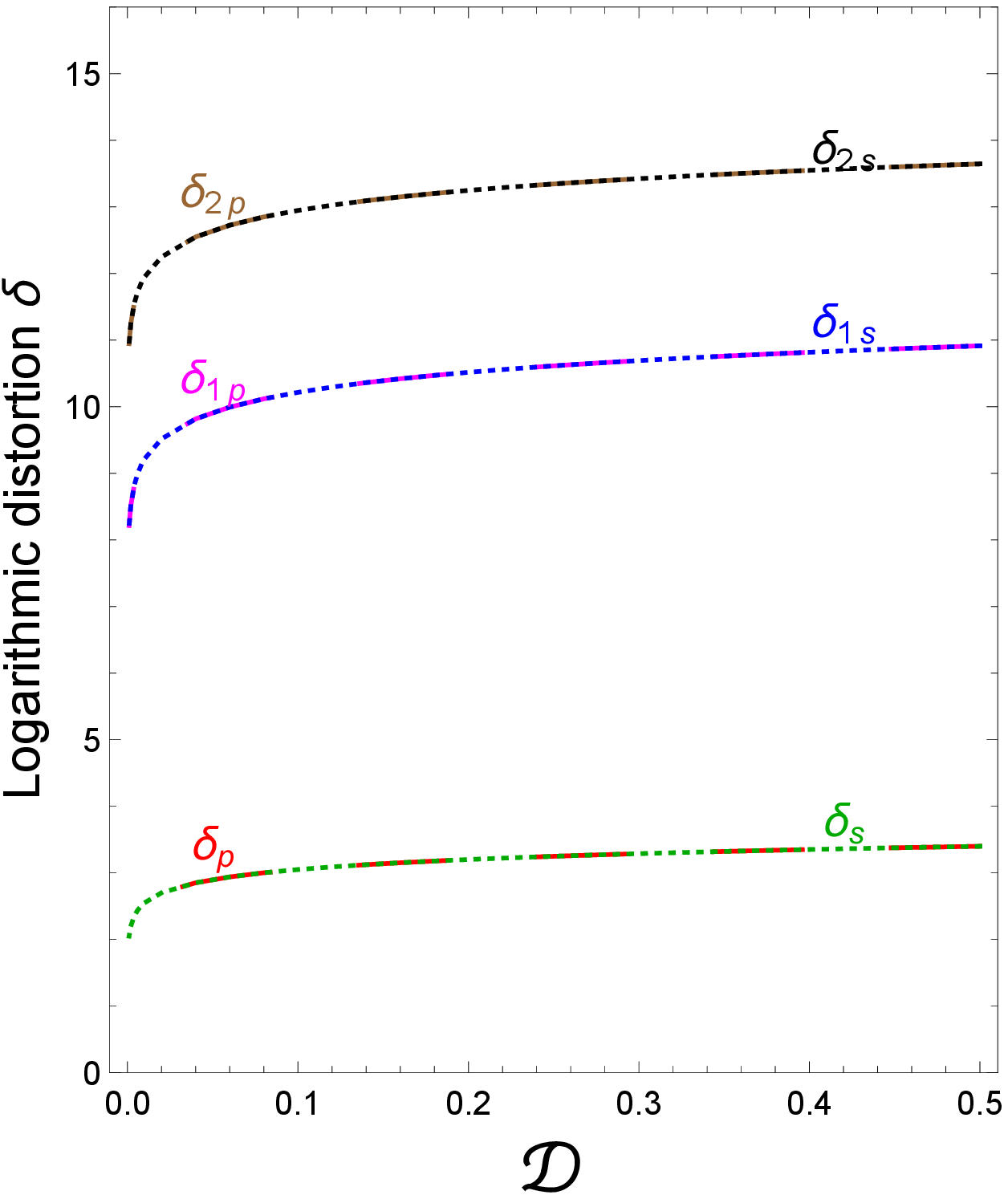} \qquad      \epsfxsize 5.5cm  \epsfbox{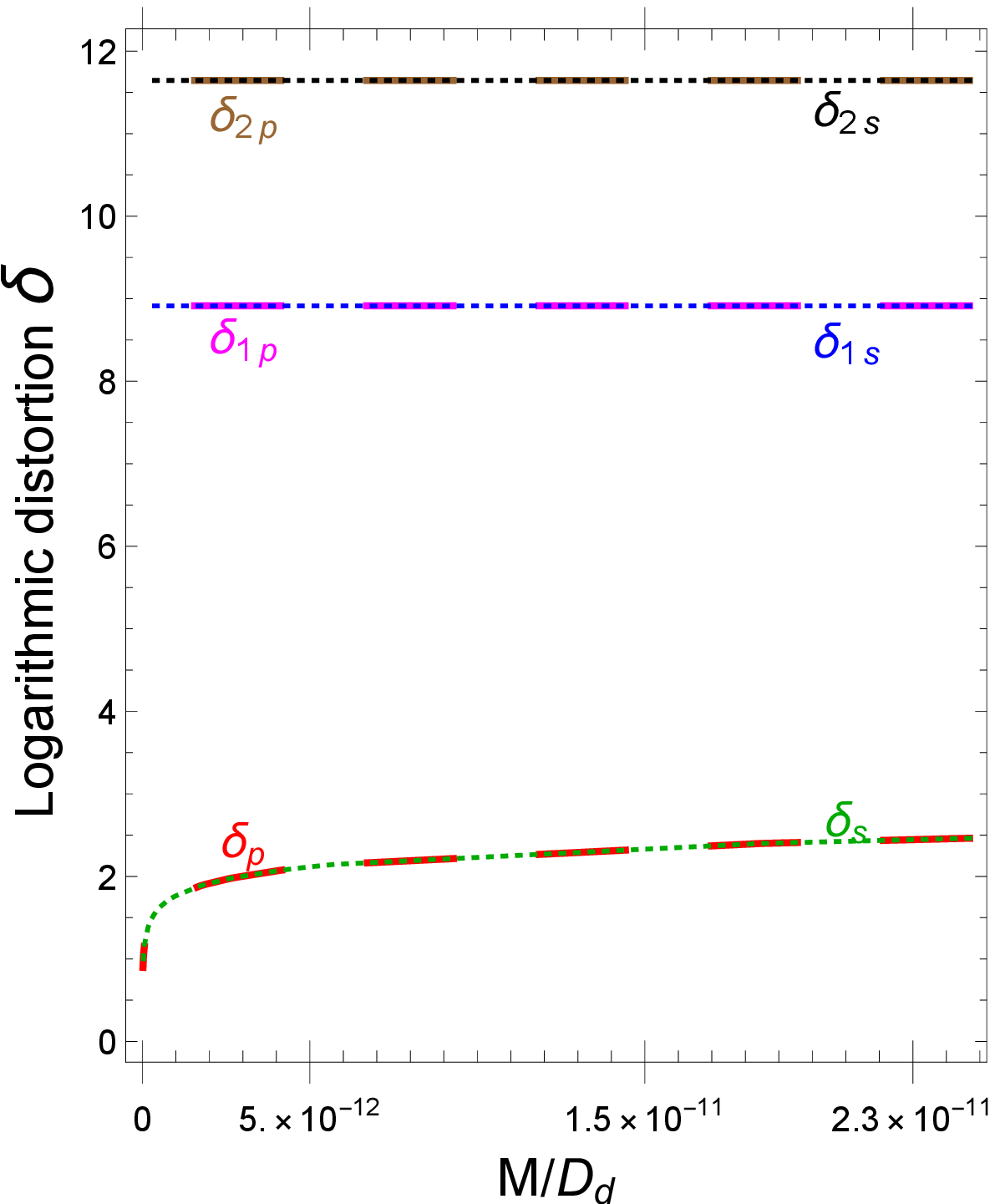}}
 \caption[ ]{
{\em Left}:  the logarithmic distortions of the primary image $\delta_p$ (red dashed), secondary image $\delta_s$ (green dotted), the first order relativistic image on primary side $\delta_{1p}$ (magenta dashed),
             the first order relativistic image on secondary side $\delta_{1s}$ (blue dotted), the second order relativistic image on primary side $\delta_{2p}$ (brown dashed), and
              the second order relativistic image on secondary side $\delta_{2s}$ (black dotted) are plotted against the angular source position $\beta$. The SMDO  at the galactic center of M87 is modeled as the    
Schwarzschild lens, which has  mass $M= 6.5 \times 10^9 M_{\odot}$ and is situated at  the distance $D_d =  16.8$ {\em Mpc}  so that  $M/D_d \approx 1.84951\times  10^{-11}$.  The dimensionless
parameter ${\cal D} = 0.005$.
{\em Middle}: the same six quantities (as for the figure on the left) are plotted against the parameter ${\cal D}$. The lens is also the same and the angular source position $\beta = 1 mas$.  
{\em Right}: the same six quantities (as for the figure on the left) are plotted against $M/D_d$ (the ratio of the mass of lens to the lens-observer distance).  
 SMDOs at the centers of 40 galaxies are modeled as    Schwarzschild lenses.  The  angular source position $ \beta = 1 mas $ and $ {\cal D} = 0.005$.  
   The colors of symbols and graphs  are kept the same for graphs to be identified.
 }
\label{fig5}
\end{figure*}

\begin{figure*}[tbh]
\centerline{
     \epsfxsize 5.5cm  \epsfbox{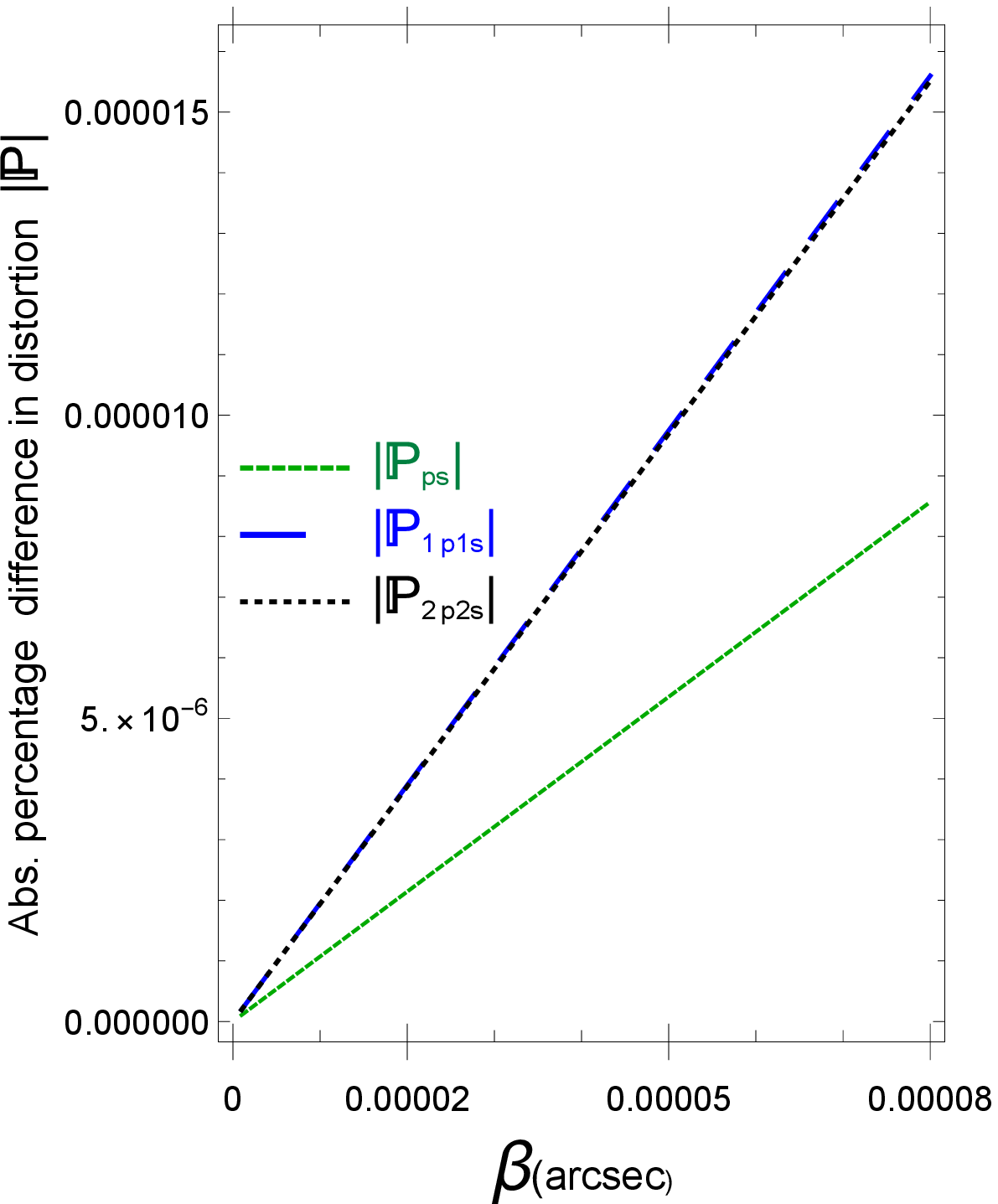}  \qquad      \epsfxsize 5.5cm  \epsfbox{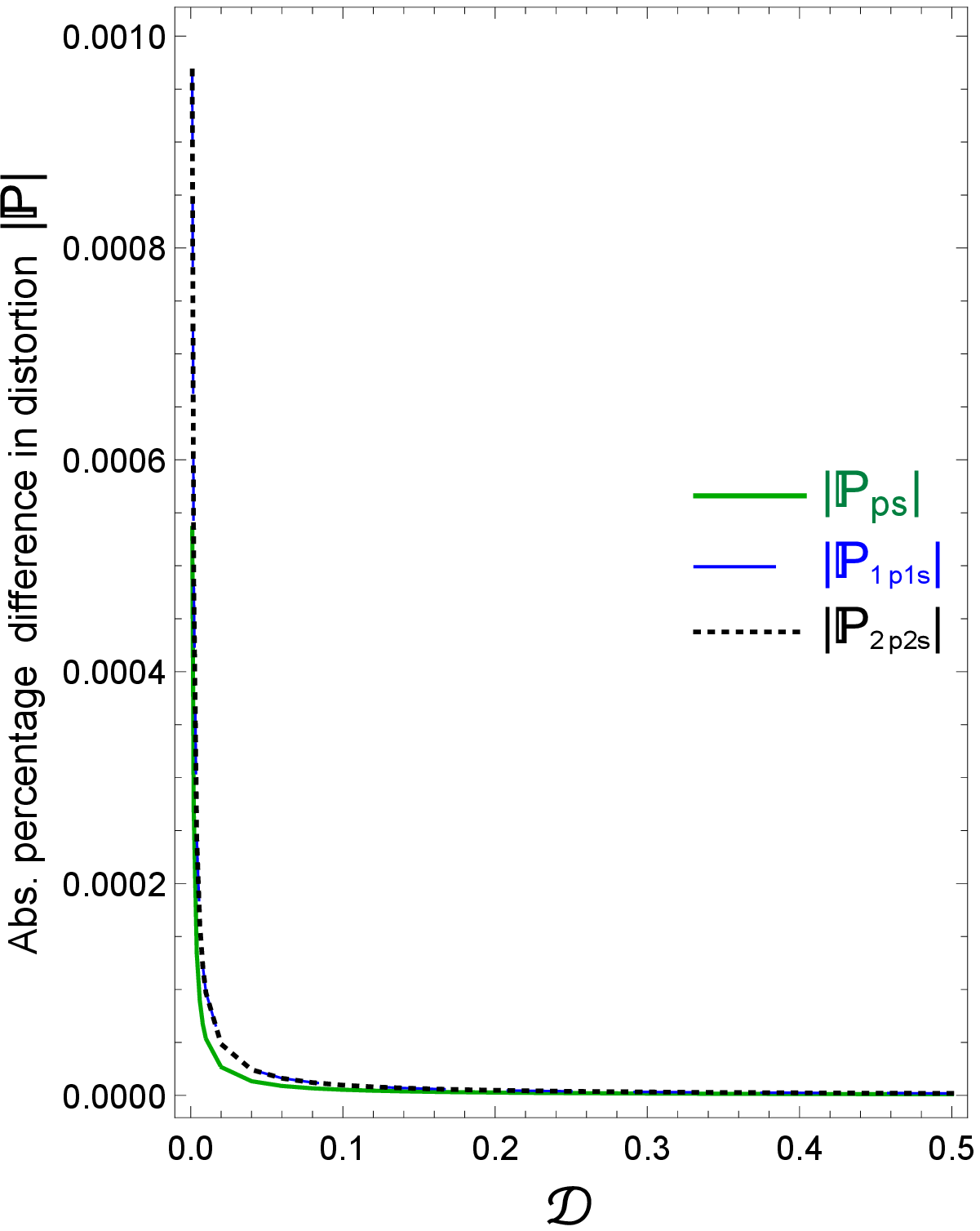} \qquad      \epsfxsize 5.5cm  \epsfbox{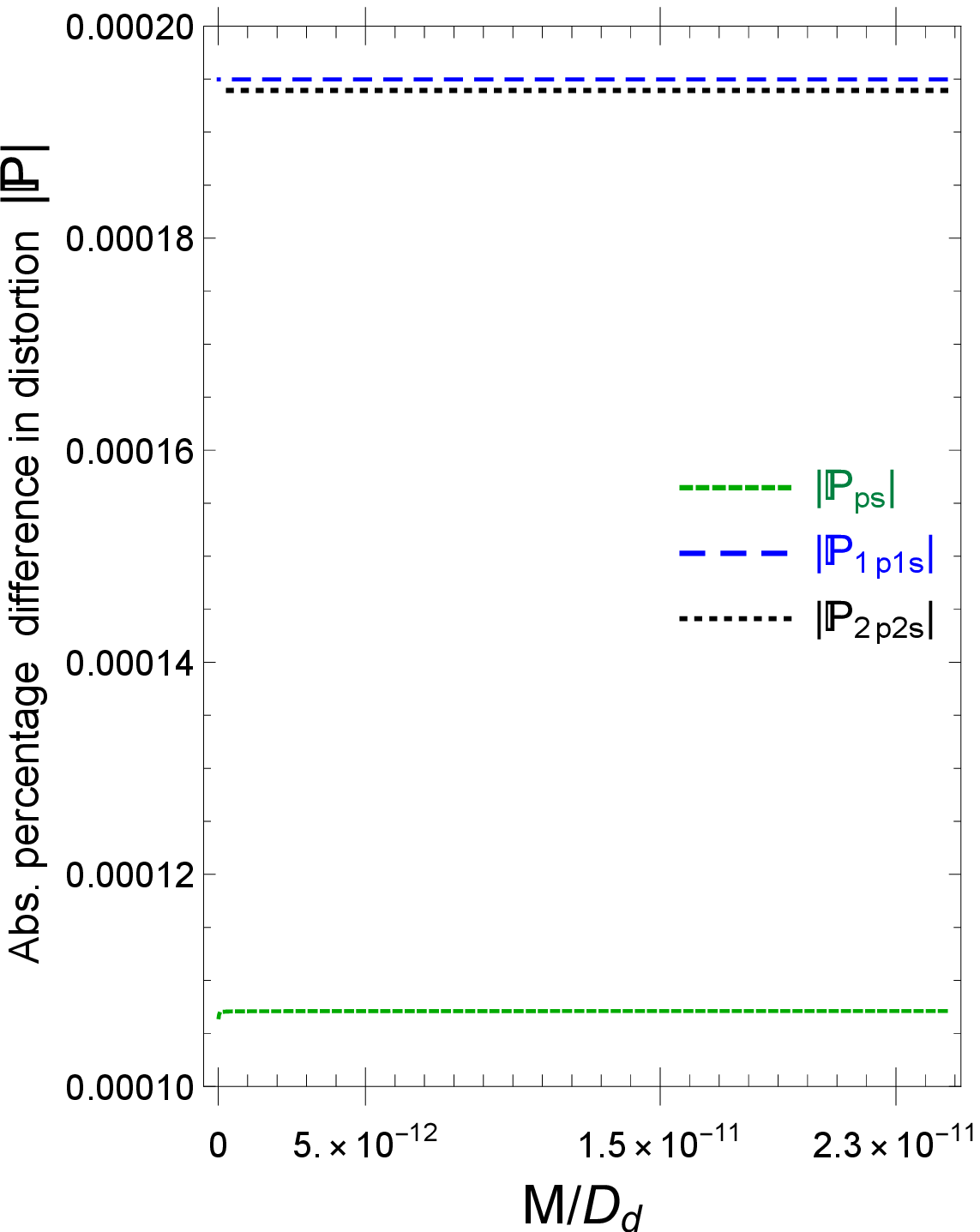}
          }
 \caption[ ]{
{\em Left}:  the absolute percentage  difference in  distortions for the primary-secondary images pair $\left|\mathbb{P}_{ps}\right|$, the first order relativistic images pair  $\left|\mathbb{P}_{1p1s}\right|$, and the second  order relativistic images pair  $\left|\mathbb{P}_{2p2s}\right|$ are plotted against the angular source position $\beta$. The MDO  at the galactic center of M87 is modeled as the   Schwarzschild lens, which has  mass $M= 6.5 \times 10^9 M_{\odot}$ and is situated at  the distance $D_d =  16.8$ {\em Mpc}  so that  $M/D_d \approx 1.84951\times  10^{-11}$.  The dimensionless parameter ${\cal D} = 0.005$.
{\em Middle}: the same three quantities (as for the figure on the left) are plotted against the parameter ${\cal D}$. The lens is also the same and the angular source position $\beta = 1 mas$.  
{\em Right}: the same three quantities (as for the figure on the left) are plotted against $M/D_d$.  SMDOs at centers of 40 galaxies are modeled as  Schwarzschild lenses. The  angular source position $\beta = 1 mas$ and ${\cal D} = 0.005$.
 }
\label{fig6}
\end{figure*}

In our paper \cite{Vir09}, we studied variations of total absolute magnifications of primary, secondary, and first- and second-order relativistic images (on the primary image side) against the angular source position $\beta$ as well as the lens mass to lens-observer distance $M/D_d$ only for three values of lens-source to the observer-source distances ratio ${\cal D}$.  However, in order to have a better knowledge of shapes of images of Schwarzschild lensing, we  now study variations of  tangential and  radial magnifications  along with  total magnifications of the primary, secondary, and first- and second-order relativistic images (on both sides of the optic axis) against $\beta$, ${\cal D}$, and $M/D_d$. These studies are at present mostly  of theoretical interest; however, these could  later be  useful to reveal important information about the lenses as well as sources.

We first model the  $M87^*$    (mass $M = 6.5 \times 10^9 M_{\odot}$ and distance  $D_d = 16.8$  {\em  Mpc }   \cite{EHT6})
as a Schwarzschild lens and study point-source gravitational lensing. In \cite{Vir09} we showed that, as opposed to cases for the primary and secondary images, total  magnifications of relativistic images of sources  with smaller values of ${\cal D}$ are higher, i.e.,  relativistic images of  sources closer to lens have higher magnifications.  Therefore, as the central thread of this paper is relativistic images, we take a small value of  ${\cal D} = 0.005$ for computations.  
For the $M87^*$ Schwarzschild lens with ${\cal D} = 0.005$, $M/D_{ds} \approx 3.68052 \times 10^{-9}$. As  the gravitational field due to the lens at the source location is  weak, the  lens equation holds good. 

We solve the gravitational lens equation $(\ref{LensEqn})$  for a large number of values of the angular source position $\beta$ and obtain positions of primary, secondary, and relativistic images of orders 1 and 2 on both sides of the optic axis. (We do not present image positions in this paper because variations of images positions with $\beta$ are studied in \cite{Vir09}.) Then we obtain their tangential, radial, and total magnifications. In Fig. 1, we plot these magnifications vs the angular source position. 
The figures in the top row show that the tangential magnifications of both primary and secondary images, represented by $\mu_{pt}$  and $|\mu_{st}|$, decrease with an increase in the value of $\beta$. The radial magnifications of primary and secondary images, respectively, increase and decrease with an increase in $\beta$. The total magnifications $\mu_p$ and $|\mu_s|$ of both images however decrease with an increase in $\beta$. The second and third rows in Fig. 1, respectively, show the variations of these magnifications with respect to $\beta$ for relativistic images of orders 1 and 2.  The graphs are qualitatively similar (but quantitatively a lot different) to the primary-secondary images pair.  The plots for relativistic images on the primary image side are qualitatively similar to those for the primary image and those on the secondary image side are similar to those for the secondary image. The tangential magnifications for all images decrease with an increase in the value of $\beta$. The radial magnifications of the primary as well as relativistic images on the primary image side increase with $\beta$; however, for secondary as well as relativistic images on the secondary image side decrease with an increase in $\beta$. The total magnifications of all images decrease with an increase in $\beta$.  All three magnifications of images are smaller for images of higher order. However, the radial magnifications of relativistic images decrease hugely faster (compared to their tangential magnifications) as the order of images increases.

We now numerically obtain partial derivatives of total magnifications  of three sets of images (two images in each set) with respect to the angular source position $\beta$ at a large number of values for $\beta$ and plot those in Fig. 2. All graphs are qualitatively similar. The partial derivatives of magnifications being negative for all images at all values of $\beta$ show, as expected, that the  total magnifications  decrease with an increase in $\beta$. However, rates of fall decrease with an increase in the value of $\beta$. The rates of fall in total magnifications decrease with the increase in the order of images. For a given value of $\beta$, the rate of fall is maximum for the images of zero order (i.e, primary-secondary images) and least for the second-order relativistic images. The rates of  fall in total magnifications  for a  pair of images of the same order are very close to each other. We will explain this in the last section.

We now again model the same SMDO (i.e., $M87^*$) as a Schwarzschild lens with the angular source position $\beta = 1 mas$. We study the behavior of tangential, radial, and total magnifications of primary, secondary, and relativistic images of orders 1 and 2 (on both sides of the optic axis) as the values of ${\cal D}$ increases from $0.001$ to $0.5$. All graphs are plotted in Fig. 3. The tangential magnifications for the primary as well as the secondary images increase with the increase in the value of ${\cal D}$ and both curves are concave down at all points. However, the behavior of tangential magnifications vs ${\cal D}$ graphs for relativistic images are quite different. The tangential magnifications of relativistic images on the primary side decrease whereas those on the secondary side increase with the increase in the value of ${\cal D}$. The $\mu_{1pt}$ and $\mu_{2pt}$ vs ${\cal D}$ graphs are concave up whereas $|\mu_{1st}|$  and $|\mu_{2st}|$ vs  ${\cal D}$ graphs are concave down everywhere. The radial magnifications of primary and secondary images, respectively, decrease and increase with the increase in the value of ${\cal D}$. The $\mu_{pr}$ vs ${\cal D}$ curve is concave up whereas $|\mu_{sr}|$ vs ${\cal D}$  curve is concave down.  However, radial magnification of relativistic images of orders 1 and 2 on the primary as well as a secondary side all decrease with an increase in the value ${\cal D}$ and all curves are concave up. The total magnifications of the primary and secondary images increase with an increase in the value of ${\cal D}$ whereas those for  relativistic images decrease with an increase in ${\cal D}$. The  graphs for the total magnifications (for primary and secondary images) vs ${\cal D}$ are concave down, whereas those for relativistic images are concave up.

We now consider SMDOs at galactic centers of 40 galaxies. The masses and distances of these SMDOs are listed in Table IV on page 14 in \cite{Vir09} (see  references therein for the source of data.) For  $SgrA^*$ and $M87^*$, we however consider the recently known values of masses and distances \cite{Abuter20,EHT6}.  As the main thread of this paper is to study relativistic images we consider small values of the angular source position $\beta = 1 mas$ and ${\cal D} = 0.005$ so that magnifications of these images are not extremely low. We compute the same magnifications (tangential, radial, and total) of the primary, secondary, and relativistic images of orders 1 and 2 on both sides of the optic axis for all 40 SMDOs as Schwarzschild lenses. In Fig. 4, we plot the magnifications vs $M/D_d$. All magnifications  of these  images (excluding the radial magnification of the primary image)  increase with the increase in the value of $M/D_d$.  The radial magnification of the primary image  decreases with the increase in the value of $M/D_d$.

We now proceed to test the distortion hypothesis for the simplest gravitational lensing, i.e., the weak field Schwarzschild lensing. Under the weak gravitational field limit with small angular source position, our lens equation reduces to the well-known lens equation\cite{Book1,Book2}:
\be
\beta = \theta -  \hat{\alpha} {\cal D} \text{,}
  \label{WeakFieldGLE}
\ee
where the Einstein bending angle  $\hat{\alpha} = 4M/r_0$ ($r_0$ is the closest distance of approach.) Solving the above lens equation gives the angular positions of the primary and secondary images of the Schwarzschild lensing,  respectively,  by \cite{Book1}:
\bea
\theta_p &=& \frac{1}{2} \left(  \beta + \sqrt{\beta^2 + 4 \theta_E^2}  \right)   \text {and} \nn\\
\theta_s &=& \frac{1}{2} \left(\beta - \sqrt{\beta^2 + 4 \theta_E^2}  \right),
 \label{ImagePositions}
\eea
where the angular radius of the Einstein ring $\theta_E = \sqrt{4 {\cal D} M/D_d}$. Using  $(\ref{ImagePositions})$  with $(\ref{MutMur})$ in $(\ref{Delta})$, we calculate  distortions of the primary image $\Delta_p$  and  the secondary image  $\Delta_s$:
\be
\Delta_p = - \Delta_s = \frac{\sqrt{\beta^2+ 16 {\cal D} \frac{M}{D_d}}}{\beta} \text{.}
\label{DeltaPS}
\ee 
Therefore, the signed sum of these distortions $\Delta_{sum} =  \Delta_p + \Delta_s$ identically vanishes. This supports our distortion hypothesis for  weak gravitational field lensing where the higher order terms in the bending angle have been neglected. Therefore, it is imperative to test our hypothesis when no such approximation is taken and relativistic images are also included. We carry out numerical computations with a very high accuracy.  Our computations are exact in the sense that we neither take weak  nor strong field approximation.

In order to obtain Fig. $\ref{fig1}$, we modeled $M87^*$  ($M/D_d \approx 1.85 \times 10^{-11}$) as a Schwarzschild lens with the lens-source to observe-distances ratio ${\cal D} = 0.005$. We obtained tangential,  radial, and total magnifications of the primary-secondary and relativistic images of the first and second orders of both sides of the optic axis for a large number of angular source positions $\beta$. Using tangential and radial magnifications of images, we now use Eq. $(\ref{delta})$ to compute logarithmic distortions of all these images. We then plot logarithmic distortions $\delta$ of six  images (primary, secondary, and relativistic images of order 1 and 2 on both sides of the optic axis) against $\beta$. The logarithmic distortions of images of the same order are so close to each other that their graphs are not resolved on plots. The logarithmic distortion of all images decreases fast near $\beta = 0$ and then  slowly with an increase in the value of $\beta$. For a given value of $\beta$, the logarithmic distortion of images increases with the order of images, i.e., the primary-secondary images are least distorted. Now to study the variation of logarithmic distortion of images with respect to the change in the value of ${\cal D}$, we use the results obtained for plotting Fig. $\ref{fig3}$.  We compute logarithmic distortions of all six images and plot against ${\cal D}$. The logarithmic distortion of all images increases fast near a very small value of ${\cal D}$ and then increases  slowly as ${\cal D}$ increases.  At last, we use results obtained for Fig. 4 to obtain logarithmic distortions of six images for 40 different SMDOs modeled as Schwarzschild lenses. We plot the logarithmic distortion vs $M/D_d$. The distortions of all images increase with the increase in the value of $M/D_d$. The graphs in Fig. $\ref{fig5}$ show that the logarithmic distortions of higher-order images are higher; however, distortions of the same-order images are too close to appear resolved. Results in Fig. $\ref{fig5}$ show that images of the same order have incredibly close values for distortions with of course opposite signs that make the sum of signed distortions close to zero.
Equation $(\ref{DeltaPS})$ shows that the absolute distortions of the primary as well as secondary images increase with the increase in the values of $M/D_d$ and ${\cal D}$; however, these decrease  with the increase in the value of $\beta$. These effects reflect in  Fig. $\ref{fig5}$ not only for the primary and secondary images, but also for relativistic images. Moreover, distortion increases with the increase in the order of images.

Now in order to analyze how close distortions of images of the same order are, we use the distortions of images and compute absolute values of the percentage difference in distortions of images of the same order. That is, using Eq. $(\ref{PercentageDiff})$, we compute  $|\mathbb{P}_{ps}|$, $|\mathbb{P}_{1p1s}|$, and $|\mathbb{P}_{2p2s}|$ for all three cases: (1)  the lens is $M87^*$, ${\cal D} = 0.005$,  and the angular source position is changing, (2) the lens is $M87^*$, $\beta = 1 mas$, and ${\cal D}$ is changing, and (3) ${\cal D} = 0.005$ and $\beta = 1 mas$, and $M/D_d$ is changing.  These three families of curves are plotted separately in Fig. $\ref{fig6}$. The graphs show that the absolute percentage differences in distortions of the images of the same order are incredibly small supporting the hypothesis that the sum of signed distortions of all images in gravitational lensing is zero. But, why are these  not  more close to zero? This appears to be due to  the approximate lens equation (as the percentage difference is higher for larger $\beta$ and smaller ${\cal D}$ where lens approximation is not good for computing   $\mathbb{P}$). Even if the  chosen distortion parameter does not work great with the hypothesis, the parameter is likely to be very useful to study and analyze images of GL.

\section{\label{sec:Intro} Discussion and Summary}
We modeled the supermassive dark object $M87^*$ as a Schwarzschild lens and first studied the variations of three magnifications (tangential $|\mu_t|$, radial $\mu_r$, and the total $|\mu|$) of images of  orders $0$, $1$, and $2$ against the angular source position $\beta$. The variations of the tangential as well as the total magnifications for all images are qualitatively similar; these decrease with increase in the angular source position. However, the most spectacular graphs are for radial magnifications. The radial magnifications of the primary as well as the relativistic images on the primary image side increase with increase in the value of $\beta$. However, those of the secondary image as well as relativistic images on the secondary image side decrease with increase in the value of $\beta$. We plotted these results in Fig. $\ref{fig1}$.  The relativistic images are usually very demagnified (compared to primary and secondary images) and therefore it is important to investigate whether their magnifications decrease very fast as the source moves away from the optic axis (i.e., $\beta$ increases). In view of this, we computed partial derivatives of total magnifications with respect to $\beta$ of six  images (primary, secondary, and relativistic images of orders $1$ and $2$ on both sides of the optic axis) for a large number of values of $\beta$. To a great surprise, magnifications of relativistic images are much more stable (compared to primary and secondary images) with respect to change in $\beta$. The computations show that images of higher orders are less unstable with change in $\beta$. These are shown in Fig. $\ref{fig2}$. Then, we studied the behavior of the three magnifications of six images as the distance parameter ${\cal D}$ increases, keeping the angular source position $\beta$ fixed. Among $18$ graphs in Fig. $\ref{fig3}$, more fascinating and nonintuitive  ones are radial magnifications of the primary-secondary pair  and tangential magnifications of relativistic images. The radial magnifications of primary images and tangential magnification of relativistic images (on the primary image side) decrease with increase in ${\cal D}$ where the radial magnifications of  secondary images as well as tangential magnifications of relativistic images on the secondary image side increase with increase in ${\cal D}$.

After the completion of magnification studies with $M87^*$ as a Schwarzschild lens, we modeled SMDOs at  the centers of $40$ galaxies (including $M87$ and the Milky Way) as Schwarzschild lenses. We computed tangential, radial, and total magnifications for  six images (primary, secondary, and relativistic images of order $1$ and $2$ on both sides of the optic axis). We kept the angular source position $\beta$ and the dimensionless distance parameter ${\cal D}$ fixed. These results are shown in Fig. $\ref{fig4}$. Excluding the radial magnification of the primary image, the variations of magnifications with increase in the value of $M/D_d$  are reasonably intuitive. All these magnifications increase with the increase in the value of $M/D_d$. However, the behavior of the radial magnifications of the primary  image with respect to the increase in the value of $M/D_d$ is  counterintuitive  and indeed magnificent.  The radial magnifications of the primary and secondary images, respectively, decrease and increase with an increase in the value of $M/D_d$ and the separation between the graphs  decreases as $M/D_d$ increases.

We hypothesized that there must exist a {\em distortion parameter} such that the signed sum of all images of  realistic and singular gravitational lensing of a source identically vanishes.  We proposed such a distortion parameter [see Eq. $(\ref{Delta})$] and, as a first step to support the hypothesis, we demonstrated that our hypothesis holds good for images of Schwarzschild lensing in weak gravitational field. However, as the weak field lensing example is not enough to say that the hypothesis is correct, we carried out numerical computations (without either weak or strong field approximations) for the images formed due to light  deflections  in weak as well as strong gravitational fields. In order to conveniently plot, we first defined a logarithmic distortion parameter  $\delta$ in  Eq. $(\ref{delta})$ and plotted it against the angular source position $\beta$, the 
dimensionless distance parameter ${\cal D}$, and the ratio of the mass to the distance of the lens $M/D_d$ for  the six images (primary, secondary, relativistic images   of orders $1$ and $2$ on  both sides of the optic axis).  We found that, for all images, the logarithmic distortion parameter decreases with an increase in $\beta$, and a decrease in ${\cal D}$ as well as $M/D_d$.  The graphs for logarithmic distortion of the images of the same order appear unresolved (see in Fig. $\ref{fig5}$) because their values are too close to each other. The sum of signed distortions of all images is extremely close to zero. 
Now, in order to see how close are the values of distortions of images of the same order, we defined and computed the absolute percentage differences of images of the same order and found these to be extremely small, and therefore these results very strongly support the distortion hypothesis. This parameter increases  with an increase in $\beta$ and a decrease in ${\cal D}$, and remains almost constant with an increase in the value of $M/D_d$. These results suggest that the nonvanishing  (though extremely small) values of the absolute percentage difference in distortions are very likely to be due to approximations involved in the lens equation and numerical computations. Thus, our hypothesis passes the Schwarzschild lensing with flying colors. However, this still remains to be tested  with many other realistic gravitational lensing.

Our hypothesis does not insist that there has to be a unique distortion parameter supporting our hypothesis. After the completion of this work, we found another distortion parameter 
\be
\Delta^* = \mu_t - \mu_r \text{.}
\ee
With this, we obtain distortions of the primary image $\Delta^*_p$ and the secondary image $\Delta^*_s$:
\be
\Delta^*_p = - \Delta^*_s =  \frac{8 {\cal D} \frac{M}{D_d} } {\beta \sqrt{16 {\cal D} \frac{M}{D_d}  + \beta^2}} \text{.}
\ee
Thus, their signed sum $\Delta^*_p + \Delta^*_s$ identically vanishes. This supports our hypothesis. However, it still remains to be tested if we include relativistic images which are formed due to light deflections in very strong gravitational fields. It is possible that our hypothesis works well only for certain types of realistic  lensing.  If so,  the examples and counterexamples could be useful  for classifying types of gravitational lensing. The theory behind the measurements of any distortion parameter satisfying our hypothesis is still to be developed. During these studies, we found some interesting relationship which are though  not much related to the present topic, is worth putting here: $\mu_{tp}+\mu_{ts} =  \mu_{rp}+\mu_{rs} = 1$ for weak field Schwarzschild lensing. Around two decades ago, we (the present author) noticed, based on numerical computations, a relation: the signed sum of (total) magnifications of images of the same order is approximately 1. This explains why the curves in Fig. $\ref{fig2}$ are too close to appear separately. It seems that the weak field  part of the  result,  $\mu_{p}+\mu_{s} = 1$,  is known in the literature. 

Despite many industrious efforts there is neither a proof nor a disproof of the  {\em weak cosmic censorship}  hypothesis which basically states that, generically, spacetime singularities of physically realistic gravitational collapse  are hidden within event horizons (see details in \cite{Vir96,VJJ97,Vir99} and references therein.)  Spacetime singularities not covered inside an event horizon are called  {\em naked singularities}. We studied gravitational lensing due to naked singularities and found that these serve as better cosmic telescopes than regular massive objects as well as black holes of the same ADM (Arnowitt-Deser-Misner) mass due to the following reasons:  Naked singularity lenses give rise to (i) higher value for the sum of absolute total magnifications of all images, (ii) smaller time delays, and (iii) smaller magnification centroid shift that enables us locate the source position better.  Thus, {\em naked (visible) singularities are not just visible to observers, these  make our universe more visible to us}. We also showed that strongly naked singularities (i.e.,  those not covered inside any photon sphere) can give rise to images of negative time delays. Naked singularities (excluding strongly naked ones) are one of the best mimickers of black holes and therefore, studies in this paper  should be extended to those naked singularities.  As  SMDOs have rotation, it is extremely important to thoroughly study magnifications and 
 distortions of images of the  Kerr lensing (including retrolensing). 

\acknowledgments
Thanks are due to F.~Eisenhauer for helpful correspondence and the anonymous referee for a careful reading of the manuscript.


\begin{references}
\bibitem{ModifiedDef} There are spacetimes whose ADM (Arnowitt-Deser-Misner) masses are zero, but cause deflection of electromagnetic as well as gravitational radiations. Due to this  reason, we have  improved the definition of gravitational lensing.
\end{references}
\end{document}